\documentclass{article}
\usepackage[english]{babel}
\usepackage[utf8]{inputenc}
\usepackage{johd}
\usepackage[table.xcdraw]{xcolor}
\usepackage{soul}
\usepackage{svg}
\usepackage{graphicx}
\usepackage{caption}
\usepackage{amssymb}
\usepackage{multirow}
\usepackage{tabularx}
\usepackage{eucal}
\usepackage{svg}
\usepackage{amsmath}
\usepackage[export]{adjustbox} 
\usepackage{colortbl}
\usepackage{caption}
\usepackage{booktabs}
\usepackage{makecell}
\usepackage{threeparttable}
\usepackage{url}
\usepackage{amsthm}
\usepackage{dsfont}
\usepackage{mathtools}
\DeclarePairedDelimiter{\ceil}{\lceil}{\rceil}
\let\oparagraph\paragraph
\renewcommand{\paragraph}[1]{\oparagraph{\rm\textbf{#1}}}

\theoremstyle{definition}

\title{Simple binning algorithm and SimDec visualization for comprehensive sensitivity analysis of complex computational models}

\author{Mariia Kozlova$^{a}$$^{*}$, Antti Ahola$^{b}$, Pamphile Roy$^{c}$, Julian Scott Yeomans$^{d}$. \\
        \small $^{a}$LUT Business School, LUT University, Lappeenranta, Finland\\
        \small $^{b}$Laboratory of Steel Structures, LUT University, Lappeenranta, Finland \\
        \small $^{c}$Austria \\
        \small $^{d}$Schulich School of Business, York University, Toronto, Canada \\\\
        \small $^{*}$Corresponding author: Mariia Kozlova; \tt{mariia.kozlova@lut.fi}
}

\date{} 

\begin{document}

\maketitle

\begin{abstract} 

Models of complex technological systems inherently contain interactions and dependencies among their input variables that affect their joint influence on the output. Such models are often computationally expensive and few sensitivity analysis methods can effectively process such complexities. Moreover, the sensitivity analysis field as a whole pays limited attention to the nature of interaction effects, whose understanding can prove to be critical for the design of safe and reliable systems. In this paper, we introduce and extensively test a simple binning approach for computing sensitivity indices and demonstrate how complementing it with the smart visualization method, simulation decomposition (SimDec), can permit important insights into the behavior of complex engineering models. The simple binning approach computes first-, second-order effects, and a combined sensitivity index, and is considerably more computationally efficient than the mainstream measure for Sobol’ indices introduced by Saltelli et al. The totality of the sensitivity analysis framework provides an efficient and intuitive way to analyze the behavior of complex systems containing interactions and dependencies.

\end{abstract}



\section{Introduction}

Engineering models of complex technological systems are characterized by a high degree of complexity and specificity. While the researchers developing them are experts in such specific mathematical approaches, as finite element methods, structural reliability and integrity modeling, and fitness-for-service (\citealp{prEN1-14, mahadevan2003ress, shen2023ress, shittu2021ress}), their professional expertise is frequently deficient in the mathematically comprehensive field of sensitivity analysis (SA) (\citealp{saltelli2019so}). Potential errors in the underlying structural analysis models can lead to failures that cause catastrophic consequences in service. To prevent such outcomes, it is essential to conduct an effective SA on those models. Furthermore, SA can also be used in model development to identify the most important parameters and their interactions and guide the process. SA can be used to explicitly test the models and exclude those parameters that do not significantly affect uncertainty in the outputs to build simplified approaches for engineering. The development of simple and intuitive SA approaches is paramount for incorporating SA into standard practice to fully complement the processes of building and analyzing a computational model (\citealp{saltelli2019so,iooss2022editorial}). 

Consequently, this paper provides several contributions to SA for assessing the behavior of complex engineering models. Firstly, we introduce a novel intuitive procedure that efficiently computes global variance-based sensitivity indices for a given dataset (Section \ref{s:si}), together with its motivation presented (Section \ref{subs:si}). Secondly, in contrast to recent mathematical developments (\citealp{lamboni2021multivariate,mara2015non,borgonovo2023total,barr2023kernel,ehre2024variance}), the new approach is shown to capture dependent inputs automatically while simultaneously revealing several interesting properties that enable examination of the relationships between dependency and interaction (Section \ref{s:corr}). Thirdly, the entirety of modern sensitivity analysis literature is focused solely on the computation of sensitivity indices (\citealp{menz2021variance,shi2023efficient,barr2023kernel,vuillod2023comparison,wang2023extended,jung2023efficient,shang2023efficient,di2023bootstrapped,jia2024improved}). In contrast, we advocate for the use of visualizations to investigate the shape of the effects in addition to their strength (Section \ref{subs:simdec}), and illustrate this with a structural reliability model that possesses nested heterogeneous interaction effects that could not be uncovered without such visual representation (Section \ref{s:simdec}). Overall, this paper introduces a simple but powerful framework for the SA of engineering models, that computes sensitivity indices to prioritize the input factors followed by a smart visualization for communicating the nature of the discovered effects. There are open-source codes in several programming languages that implement the entire framework to accompany the article.     

\subsection{Sensitivity indices computation}
\label{subs:si}

Understanding the behavior of computational models forms the basis for design and decision-making. The decision-maker obtains information on: what sources of uncertainties are most crucial and might require protection against; which actionable parameters make the most difference and thus represent the perfect levers for managing the system; and, which model parameters create the most noise and require clarification to obtain a clearer representation of the system being modeled.

\begin{figure}[H]
\centering
\includegraphics[scale=0.4]{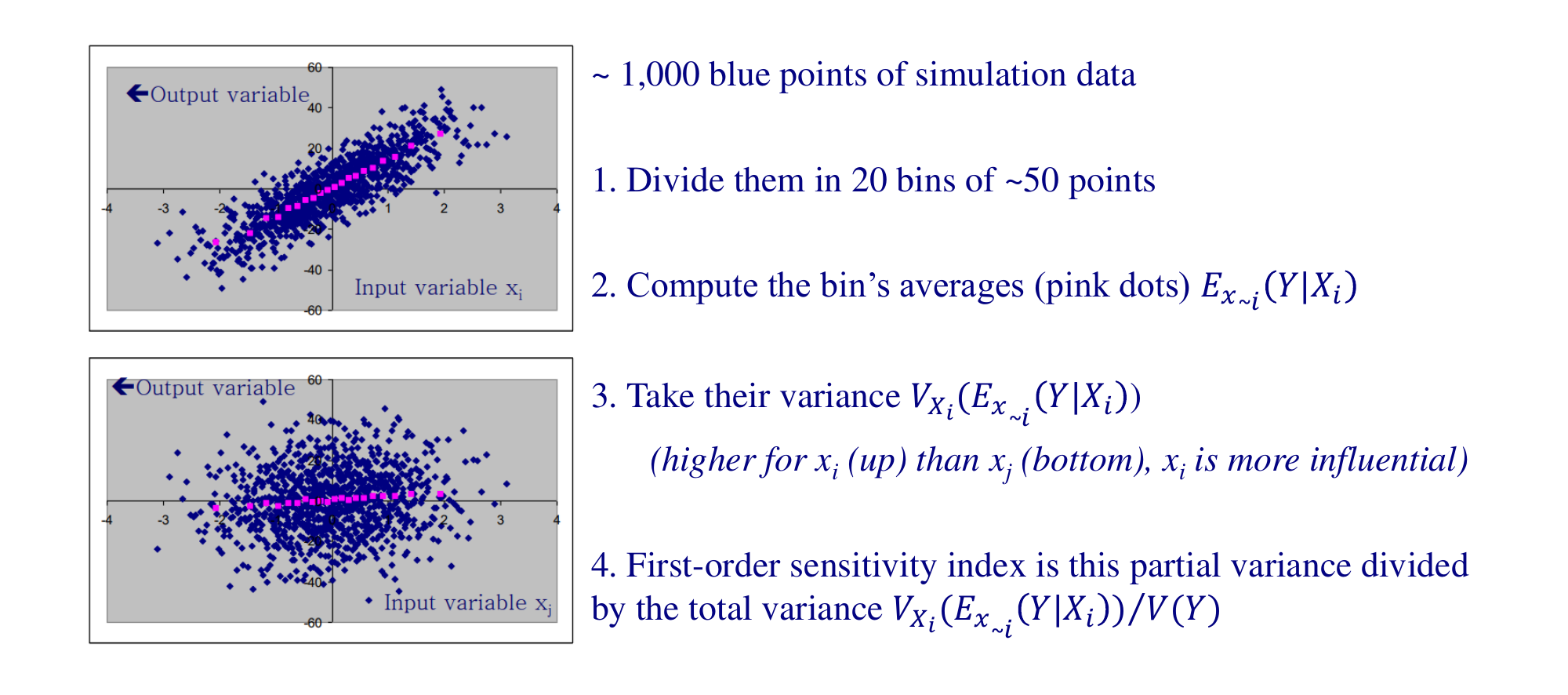}
\caption{\label{fig:pres}Conceptual representation of the sensitivity indices, adapted from (\citealp{Saltelli2023}).}
\end{figure}

The sole purpose of the scientific field of SA is to identify which input variables affect the output(s) the most in the computational models (\citealp{saltelli2008global}). For educational purposes, the conceptual idea behind the sensitivity indices is often explained using scatter plots (\citealp{Caers2018,Saltelli2023}) as depicted in Figure \ref{fig:pres}, borrowed from one of the talks devoted to SA (\citealp{Saltelli2023}).

Figure \ref{fig:pres} displays the intuition behind the variance-based sensitivity indices: namely, bin the $X$, compute the average of $Y$ in every bin, compute the variance of these average values of $Y$, and divide the resulting conditional variance to the total variance of $Y$ to obtain the sensitivity index value. Quite unexpectedly, however, no common methods have previously adopted this logic in computing sensitivity indices. Among the widely used methods for computing sensitivity indices: some involve game theory (\citealp{shapley1953value}) and some Fourier analysis (FAST) (\citealp{cukier1973study}); the classic Sobol' indices (\citealp{sobol1993sensitivity, homma1996importance,saltelli2002making,saltelli2010variance}) require juggling multiple simulation matrices; some methods employ polynomial chaos expansion (\citealp{sudret2008global,crestaux2009polynomial}); there is a so-called moment-independent measure which is based on quantifying the shift in the distribution of the output caused by an input (\citealp{borgonovo2007new}); some are based on derivatives (\citealp{sobol2010derivative, kucherenko2016derivative}); there is a variogram-based method (VARS) (\citealp{razavi2016new}); while the most recent approach utilizes discrepancy measures (\citealp{puy2022discrepancy}) to quantify the difference between the cumulative probability distributions of simulated data and its running averages. Modern research on sensitivity analysis is actively developing and predominantly targets an increase in the computational efficiency of these various methods (\citealp{iooss2022editorial,saltelli2021sensitivity}). 

A simple method that follows the logic of Figure \ref{fig:pres} was independently introduced by \cite{marzban2016conceptual} as a \textit{conceptual implementation of Sobol' indices} and by \cite{kucherenko2017sobol} as the \textit{double loop rendering method or DLR}. Our focus is drawn to this method because:
        \begin{itemize}
            \item it is intuitive and works precisely as sensitivity indices are conceptually introduced  (Figure \ref{fig:pres});
            \item it is more computationally efficient and more accurate than other methods for estimating Sobol' indices (\citealp{kucherenko2017different});
            \item it works with a given dataset. As \cite{plischke2012compute} notes, for such methods: (i) no access to the model is required; (ii) they can work on measured data; and as a result, (iii) the method is indifferent to distribution types and sampling methods. Furthermore, it enables smoother passing of information between the model and sensitivity analysis, if they are performed in different software or by different people.     
        \end{itemize}
This paper extends this concept to explicitly capture the second-order effects. Such an approach has not been considered previously, as \cite{marzban2016conceptual} algorithm determined the first-order effects, exclusively, while \cite{kucherenko2017sobol} estimated the first-order and the total effects. This modified new approach, referred to as the \textit{Simple binning approach}, introduces additional benefits:
        \begin{itemize}
            \item the choice of the number of bins is no longer the user's responsibility, it is automated based on experiments by \cite{marzban2016conceptual};
            \item it works with dependent inputs (although developments that allow sensitivity analysis of models with dependent input variables exist (\citealp{da2015global,lamboni2021multivariate,mara2015non,borgonovo2023total,ehre2024variance}), majority of them require certain additional transformations, whereas the simple binning approach captures dependent inputs "as is" (\citealp{kucherenko2017sobol}) and preserves the conservation property);
            \item the measure for the second-order effect allows studying the behavior of dependent inputs that affect the output interactively, revealing dependency-interaction intertwining; 
            \item it works with categorical/discrete variables; 
            \item it is resistant to different numbers of observations in bins due to the usage of weighted variance;
            \item as a result, it can work with empirical datasets, or 
            \item it can be used to quantify the sensitivity of the output to intermediate outputs in simulation models.
        \end{itemize}

\subsection{Importance of visualization}
\label{subs:simdec}

The modern practice of sensitivity analysis systematically stops at computing sensitivity indices (\citealp{menz2021variance,shi2023efficient,barr2023kernel,vuillod2023comparison,wang2023extended,jung2023efficient,shang2023efficient,di2023bootstrapped,jia2024improved}) and does not take an additional step of investigating the shape of the effects in a model. Model interactions may arise in many different forms (\citealp{kozlova2024uncovering}). In one model, they might be the result of multiplication in which the effect of one input variable linearly increases in the higher values of another input variable. In another model, the effect of the variable might be strong in a certain region of another input variable, but negligible in the others. The direction of the influence of one input factor on the output might be flipped in different regions of another input variable. In all of these cases, the sensitivity index will show a positive second-order effect of a certain degree, but no more. Namely, no indication of the \textbf{\textit{type}} of interaction effect can be obtained through sensitivity indices alone. 

However, explicit knowledge of the nature of interaction effects is crucial for effective decision-making and engineering design and, thus, visualization of the effects is a "must-have" practice for computational model analysis. Heat maps (\citealp{pleil2011heat,owen2019acas}) and response surfaces (\citealp{myers2016response}) are inherently three-dimensional visualizations that can portray the joint effect of a pair of input variables on the model output. However, these visualization types cannot handle simultaneous variations of multiple inputs and, thus, fail to depict higher-order interactions. 

These visualization limitations are overcome by the \textit{simulation decomposition} (SimDec) approach (\citealp{kozlova2022monte}). SimDec partitions the output probability or frequency distribution into sub-distributions comprised of combinations of specific regions of influential input variables, thereby exposing the nature of interaction effects. The SimDec approach has demonstrated value in multiple applications (\citealp{kozlova2019multi, deviatkin2021simulation,liu2022analysis}). In this paper, it will be shown that the concurrent employment of the simple binning approach to identify the most influential input variables with SimDec to display the nature of those effects results in a simple-to-implement, intuitive, and sophisticated framework for conducting SA.

\section{Simple binning approach to sensitivity indices}
\label{s:si}

\subsection{Existing approach for first-order effects}
\label{subs:marzban}
\cite{marzban2016conceptual} and \cite{kucherenko2017sobol} introduced a conceptual implementation to Sobol' first-order indices. Unlike the specific design of experiments normally needed for computing the estimators of Sobol' indices (several sets of Monte Carlo simulation data with different input variable(s) fixed) (\citealp{saltelli2002making,saltelli2010variance}), their approach requires only a single $N_{observations}*K_{inputs}$ dataset generated from one Monte Carlo simulation. The approach involves binning along the $X$’s range, computing averages of $Y$s within every bin, and then taking the variance of those averages (as shown in Figure \ref{fig:pres}). The first-order sensitivity index is then computed as the conditional variance divided by the variance of $Y$, according to the formal description of Sobol' indices (\citealp{sobol2001global}):

\begin{equation} \label{eq:first_order}
\ S_{x_i} = \frac{\operatorname{Var}(\mathbb{E}(Y \mid X_i))}{\operatorname{Var}(Y)} 
\end{equation}

\cite{kucherenko2017sobol} also provide an algorithm for the estimation of the total effects (the effects of individual input variables containing all their interactions) and demonstrate that the method is able to handle dependent inputs. \cite{kucherenko2017different} subsequently ran multiple tests that conclusively demonstrate that this method is more computationally efficient and more accurate than other estimators.

To estimate the number of bins, \citet{kucherenko2017different} propose a "rule of thumb" $\sim\sqrt{N_{obs}}$, while \citet{marzban2016conceptual} define the optimal number of bins for different \(N_{obs}\) and \(K_{inputs}\), which results in more conservative estimates for this parameter, Table \ref{table:binning}.

\begin{table}[ht]
\scriptsize
\caption{Recommendations for the choice of the number of bins by \citet{kucherenko2017different} (second column) and \citet{marzban2016conceptual} (other columns).}
\centering
\label{table:binning}
\begin{threeparttable}
\begin{tabular}{r>{\raggedleft\arraybackslash}p{2cm}r>{\raggedleft\arraybackslash}p{2cm}>{\raggedleft\arraybackslash}p{2cm}>{\raggedleft\arraybackslash}p{2cm}} 
\toprule

                               &    & \multicolumn{3}{c}{Number of input variables, $K_\text{inputs}$} \\
                                     \cmidrule{3-5}
Number of observations, $N_\text{obs}$  &  $\sqrt{N_\text{obs}}$  & 3               & 6              & 12                \\
\midrule

1000                         &  32   & 10              & 10             & 10               \\
2500                         &  50   & 25              & 10             & 10               \\
5000                         &  71   & 50              & 10             & 10               \\
7500                         &  87   & 50              & 25             & 10               \\
10000                        &  100  & 50              & 50             & 10               \\
25000                        &  158  & 100             & 50             & 25               \\
50000                        &  224  & 100             & 50             & 50               \\ \hline
\multirow{2}{*}{Recommended by}               & (\citealp{kucherenko2017different}) & \multicolumn{3}{c}{(\citealp{marzban2016conceptual})} \\

\bottomrule
\end{tabular}
\end{threeparttable}
\end{table}
A smaller number of bins results in less noisy estimates for sensitivity indices. Therefore, the recommendation of (\citealp{marzban2016conceptual}) is followed for the simple binning code implementation and automated with linear fitting clamped at 10. 

\begin{equation} \label{eq:n_bins}
    \begin{array}{l}
m_{bins} = \ceil{36 - 2.7* K_{inputs} + (0.0017 - 0.00008 * K_{inputs}) * N_{obs}}\\
M_{bins} = 
        \begin{cases}
          m_{bins} & \text{if $m_{bins}>30$}\\
          10 & \text{otherwise}\\
        \end{cases}  
    \end{array}
\end{equation}


\subsection{Extension to second-order effects}

\begin{figure}[H]
\centering
\includegraphics[scale=0.5]{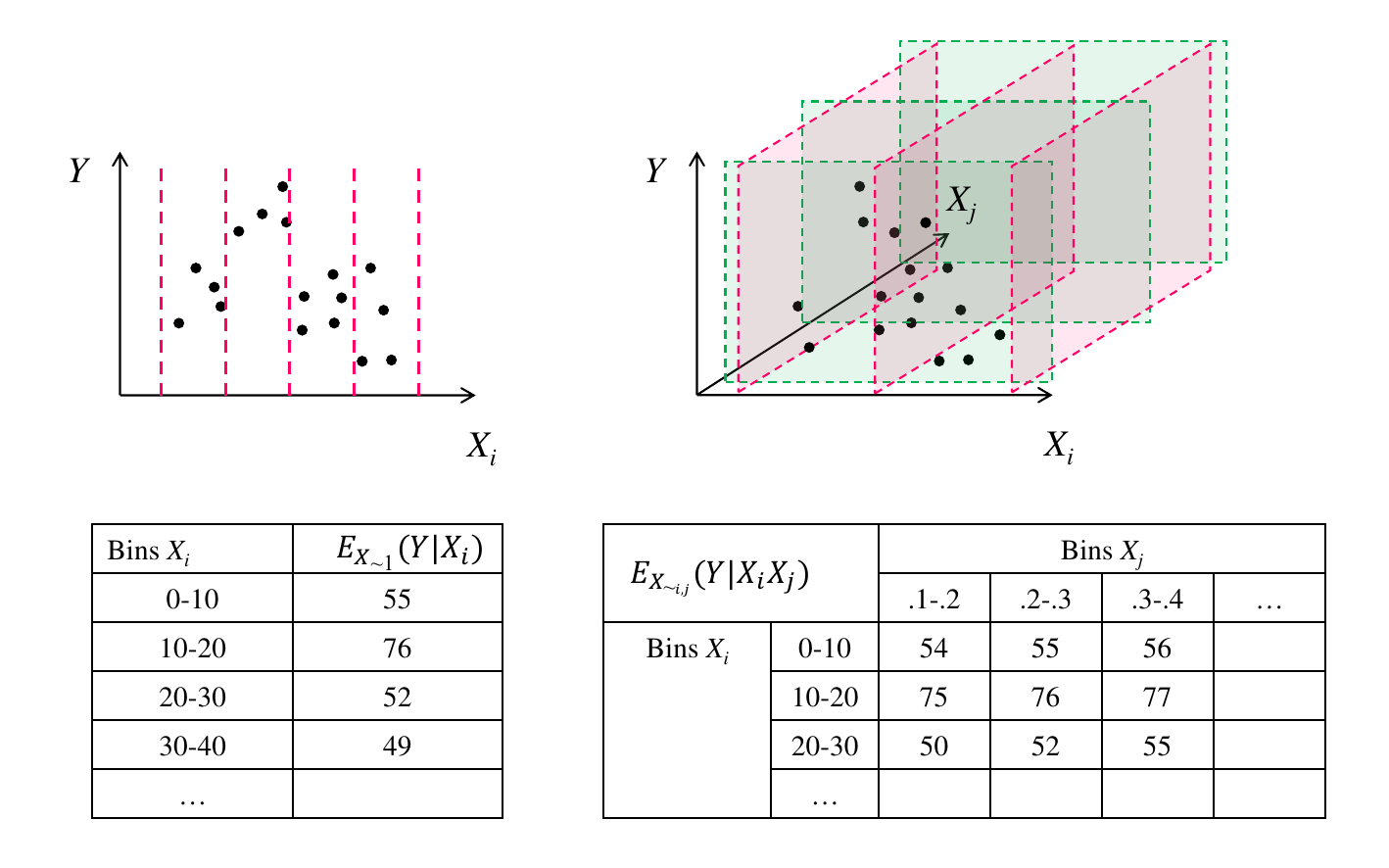}
\caption{\label{fig:binning}Simple binning approach for first-order indices (left) and second-order indices (right).}
\end{figure}

Following on from the first-order index calculation approach, we extend the binning idea into the computation of second-order effects. This implies that instead of binning a line defined by a single input, we must now bin a two-dimensional area defined by two inputs. Figure \ref{fig:binning}. 

The calculation steps follow the same procedure for computing averages in the bins and their variance, except we employ a weighted variance to account for possibly unequal numbers of observations in bins relevant to categorical variables with different frequencies for different categories, and for second-order effects:

\begin{equation} \label{eq:weight_var}
\operatorname{Var}_w(Y_{X_i}) = \frac{\sum_{b=1}^{N_\text{bins}}N_{\text{obs}_b}(\overline{Y_b}-\overline{Y})^2}{\sum_{b=1}^{N_\text{bins}}N_{\text{obs}_b}}.
\end{equation}

\noindent where the binning happens over $X_i$ and the averages are taken of $Y$ values as illustrated in Figure \ref{fig:binning}.

The second-order effect is defined in the classic way (\citealp{saltelli2002making}).

\begin{equation} \label{eq:2}
\ S_{{X_i}{X_j}} = \frac{V(E(Y|X_iX_j)}{V(Y)} -  S_{X_i} - S_{X_j}
\end{equation}
or
\begin{equation} \label{eq:AB}
\ S_{{X_i}{X_j}} = \frac{V(E(Y|X_iX_j)}{V(Y)} -  \frac{V(E(Y|X_i)}{V(Y)} - \frac{V(E(Y|X_j)}{V(Y)}
\end{equation}

The number of bins is defined based on the optimal number of observations per bin. For the first-order effects, the number of bins is a linear approximation of the optimal experimental number of bins as a function of the number of simulation runs and the number of inputs, eq. \ref{eq:n_bins}. For the calculation of second-order effects, to preserve approximately the same number of observations in two-dimensional bins, the range of each input variable is broken down into a number of intervals $M_{bins_{soe}}$ equal to a square root (rounded to the nearest integer) of the previously computed number of bins for the first-order effects.

\begin{equation} \label{eq:n_bins_soe}
M_{bins_{soe}} = \ceil{\sqrt{M_{bins}}}
\end{equation}

The same number of bins should be applied to all terms in Equation \ref{eq:AB}. For example, a stylized scheme in Figure \ref{fig:binning} depicts four bins for $X_i$ for the first-order effect, which translates into two bins for $X_i$ and two for $X_j$ for second-order effects, thus,  the bin size remains the same. 

An aggregate or combined sensitivity index (also known as a closed index) can now be calculated which, for each input variable, sums up its first-order effect and halves of second-order indices with all other input variables:

\begin{equation} \label{eq:comb}
\ S_{\text{combined}_{X_i}} = S_{X_i} + \sum_{\substack{j=1\\j\neq i}}^{K_\text{inputs}} \frac{S_{{X_i}{X_j}}}{2}
\end{equation}

This combined sensitivity index aggregates the individual and the interaction effects of each input variable. The sum of combined indices for all model input variables would equal $1$ if there are no effects higher than second-order in the model and no unaccounted variation. However, due to the approximation nature of the method, potentially noisy results due to binning (\citealp{marzban2016conceptual}), and the imperfections of sampling, the sum of combined sensitivity indices can slightly deviate from $1$ even in the absence of higher-order effects. Irrespectively, we advocate for the usage of this approach in combination with SimDec to determine the selection of the prime input variables for decomposition. As long as the ranking of inputs' importance is preserved, noise does not prove to be an impediment.

\section{Testing the simple binning approach}
\label{s:test}

\subsection{Capturing interactions -- The toy model}

To demonstrate the efficacy of the simple binning approach in capturing second-order effects, we examine the simple toy model presented in the seminal sensitivity analysis textbook by \cite{saltelli2004sensitivity}. At the heart of the toy model are summations and multiplications which are basic operations in any computational model. The model equation and the input distributions are as follows:

\begin{equation} \label{eq:4}
\ Y = C_s P_s + C_t P_t + C_j P _j \\
\end{equation}
where \(P_s \sim\ N(0; 4)\), \(P_t \sim\ N(0; 2)\), \(P_j \sim\ N(0; 1)\), \(C_s \sim\ N(250; 200)\), \(C_t \sim\ N(400; 300)\), and \(C_j \sim\ N(500; 400)\).

The mainstream evaluation of Sobol' sensitivity indices is based on the proposed by Saltelli et al. estimator (\citealp{saltelli2002making, saltelli2010variance}) and open-sourced in the Python library for sensitivity analysis SALib. A side-by-side comparison of Sobol' sensitivity index results obtained by the Saltelli implementation (\citealp{saltelli2002making}) as reported in (\citealp{saltelli2004sensitivity})\ and the new simple binning method are presented in Table \ref{table:portfolio}.

\begin{table}[H]
\centering 
\caption{\label{table:portfolio} Comparison of the Sobol' indices obtained by the Saltelli implementation (\citealp{saltelli2002making}) and simple binning approach on a toy model.}
\begin{tabular}{lccc}
\hline
Effect                      & Saltelli implementation (\citealp{saltelli2002making})  & Simple binning method & Delta \\ \hline
\multicolumn{4}{l}{First-order effects}                                           \\
$P_s$                          & 36 \%  & 35 \%                     & $-1$ \% \\
$C_s$                          & 0 \%   & 1 \%                      & 1 \%  \\
$P_t$                          & 22 \%  & 20 \%                     & $-2$ \% \\
$C_t$                          & 0 \%   & 1 \%                      & 1 \%  \\
$P_j$                          & 8 \%   & 8 \%                      & 0 \%  \\
$C_j$                          & 0 \%   & 2 \%                      & 2 \%  \\
Sum of first-order effects     & 66 \%  & 67 \%                        & 1 \%  \\ \hline
\multicolumn{4}{l}{Second-order effects (of selected pairs of variables)}                                          \\
$P_sC_s$                        & 18 \%  & 16 \%                     & $-2$ \% \\
$P_tC_t$                        & 11 \%  & 10 \%                     & $-1$ \% \\
$P_jC_j$                        & 5 \%   & 6 \%                      & 1 \%  \\
Sum of second-order effects     & 34 \%  & 32 \%                     & $-2$ \% \\ \hline
Sum of all effects              & 100 \% & 99 \%                     & $-1$ \% \\ \hline
Model evaluations               & 21000  & 1000                      &      \\ \hline
\end{tabular}
\end{table}

Table \ref{table:portfolio} demonstrates that the simple binning method produces meaningful sensitivity indices close to the Saltelli implementation of the Sobol' indices (\citealp{saltelli2002making}). However, the simple binning method used only 1000 model evaluations, while 21000 data points were required to compute the Saltelli measure. The number of bins used was ten for first-order effects and three for second-order effects.

\begin{figure}[H]
\centering
\includegraphics[scale=0.7]{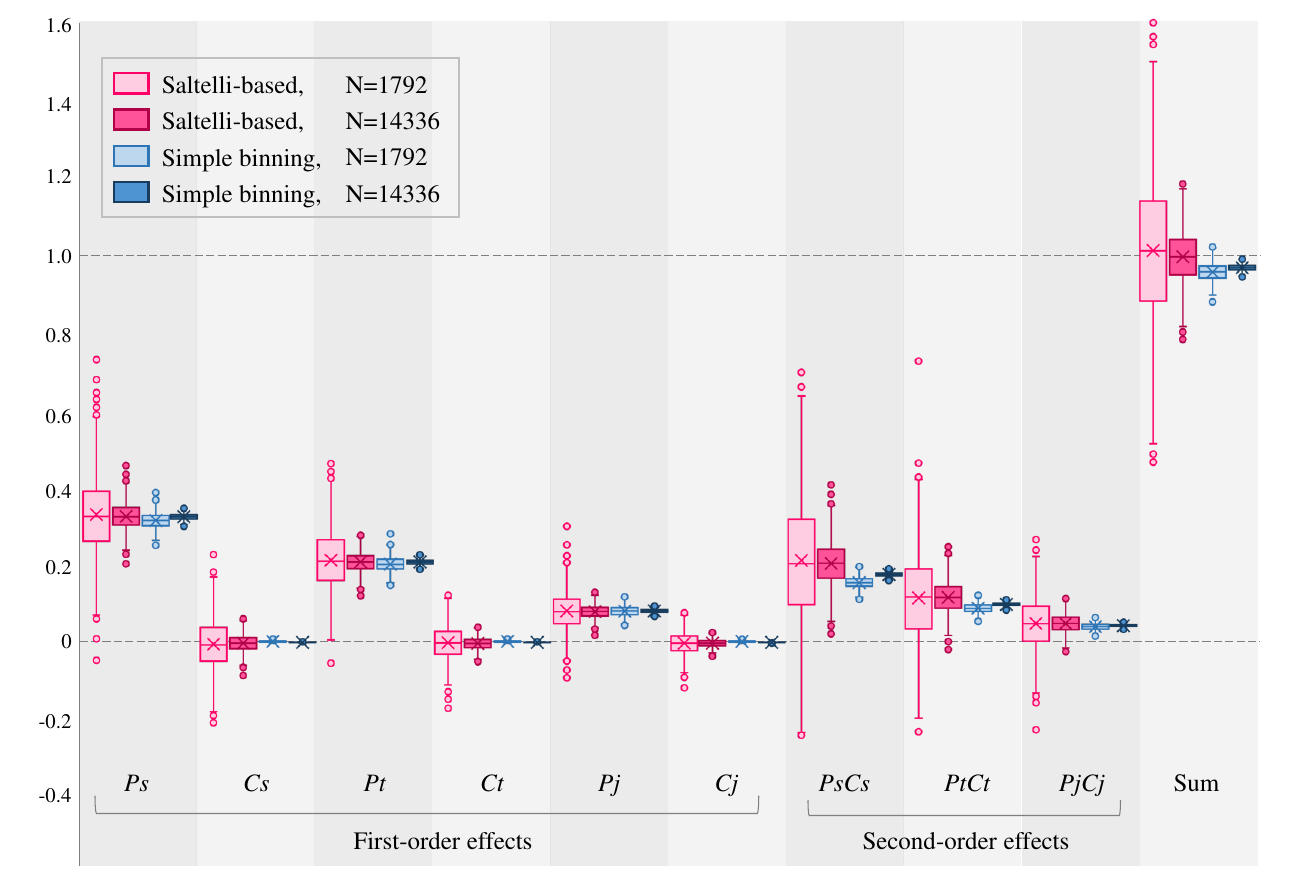}
\caption{\label{fig:port_norm}Sobol' indices for the toy model obtained with the Saltelli implementation (\citealp{saltelli2002making}) based on simple random sampling of sizes $N = 128*(2*6+2)=1792$ and $N=1024*(2*6+2)=14336$, and corresponding indices obtained with the simple binning approach.}
\end{figure}

The indices produced by both approaches in Table \ref{table:portfolio} are based on limited model evaluations with simple random sampling. Because different simulation runs can generate to some degree different datasets, we repeated the exercise 1000 times to ascertain how stable the index estimations were. Figure \ref{fig:port_norm} provides a comparison of the results obtained with the Saltelli implementation and the simple binning method for two sampling sizes. The sample size for the Saltelli implementation is dependent on the number of inputs analyzed $K$ and the specificity of the approach of fixing factors in the simulation matrices following the equation $N = n (2K+2)$, where $n$ should be a power of two. The two sample sizes chosen are 
$N = 128(2*6+2)=1792$ and $N=1024(2*6+2)=14336$. The simple binning approach produces more accurate estimates than the Saltelli implementation, even if the larger sample for the latter is contrasted with the smaller sample for the former.

\begin{figure}[H]
\centering
\includegraphics[scale=0.7]{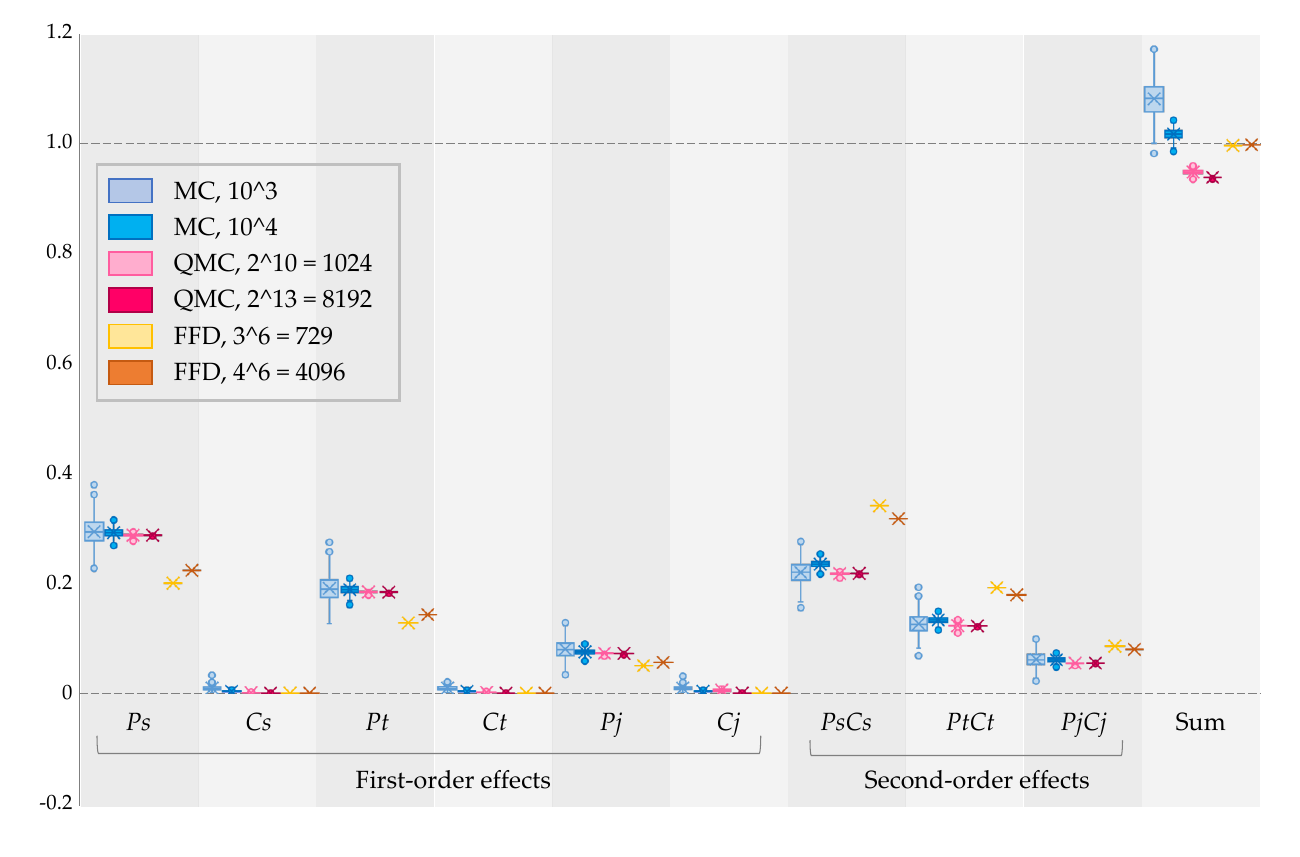}
\caption{\label{fig:port_uniform}Toy model sensitivity indices estimation by the simple binning approach with simple random sampling Monte Carlo simulation (MC), quasi-random sampling or quasi-Monte Carlo (QMC), and full factorial design (FFD) sampling.}
\end{figure}

In addition, we compared how the simple binning approach performs with (i) simple random sampling (\citealp{olken1986simple,singh2003simple}), which is often used for Monte Carlo simulation, (ii) quasi-random sampling (\citealp{sobol1967distribution,Owen:2023practical}) with the Sobol' low discrepancy sequence, which is reportedly a more efficient sampling strategy that fills the space more uniformly (\citealp{kucherenko2015exploring}), and (iii) full factorial design, frequently used in engineering contexts (\citealp{alidoosti2013electrical,suard2013sensitivity,tong2006refinement}), Figure \ref{fig:port_uniform}. For this experiment, we change the distribution of inputs from normal to uniform, selecting as minimum and maximum values $\pm$ two standard deviations of the corresponding normal distributions.

From Figure \ref{fig:port_uniform}, one can observe that simple random sampling results in noisier estimates and that this noise increases substantially the smaller the sample. Full factorial design sampling results in deterministic estimates for sensitivity indices, but the simple binning approach performed on such a sample underestimates first-order indices and overestimates second-order effects. Quasi-Monte Carlo (QMC), done with scrambling, gives more accurate and clear estimates than other sampling strategies, even with a smaller sample, and, thus, is recommended for sensitivity indices computation. QMC is easy to implement and its coded functions are available in Python (\citealp{scipy2023}), R (\citealp{chalabi2023}), Julia (\citealp{robbe2018}), and Matlab (\citealp{mathworks2013}). However, when using any of these packages, it is important to have the first point in the sequence sampled for uniformly distributed inputs, otherwise, its performance drops (\citealp{owen2020dropping}).

\subsection{Capturing cyclic behavior -- The Ishigami function}

The Ishigami function, equation \ref{eq:ishigami}, is one of the benchmark functions often used in sensitivity analysis studies for the validation of different methods, since analytical solutions for its sensitivity indices can be readily determined.

\begin{equation} \label{eq:ishigami}
\ Y = \sin(X_1) + a \sin(X_2)^2 + b  X_3^4  \sin(X_1) \\
\end{equation}

\begin{figure}[H]
\centering
\includegraphics[scale=0.55]{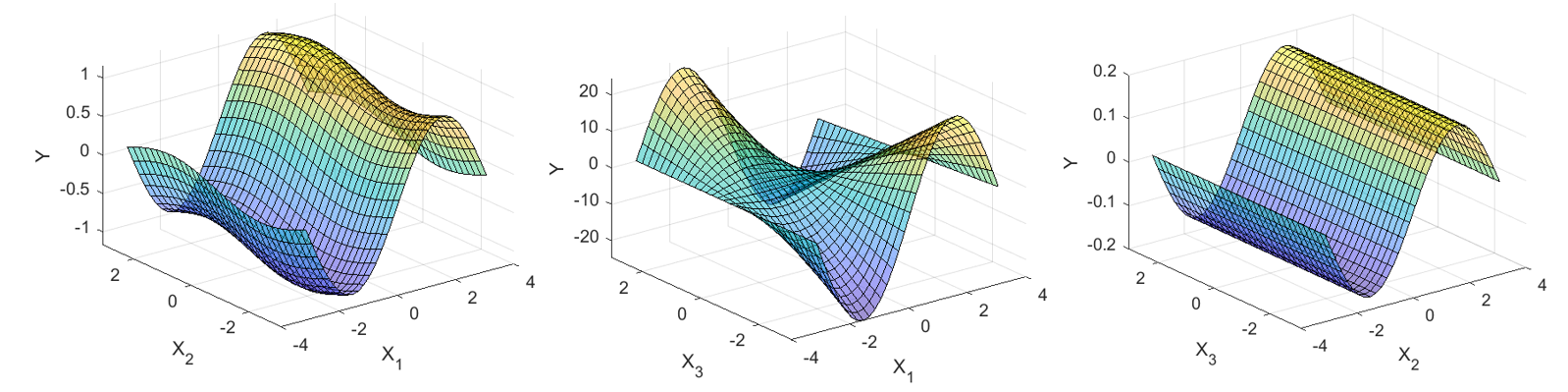}
\caption{\label{fig:ishigami_surf}Ishigami function, equation \ref{eq:ishigami}, with $a = 7$ and $b = 0.1$ (the third factor is fixed at 0 on each plot).}
\end{figure}

The function is periodic in nature, Figure \ref{fig:ishigami_surf}, which represents a significant challenge for approximate methods (\citealp{ziehn2009gui}). 

\begin{figure}[H]
\centering
\includegraphics[scale=0.7]{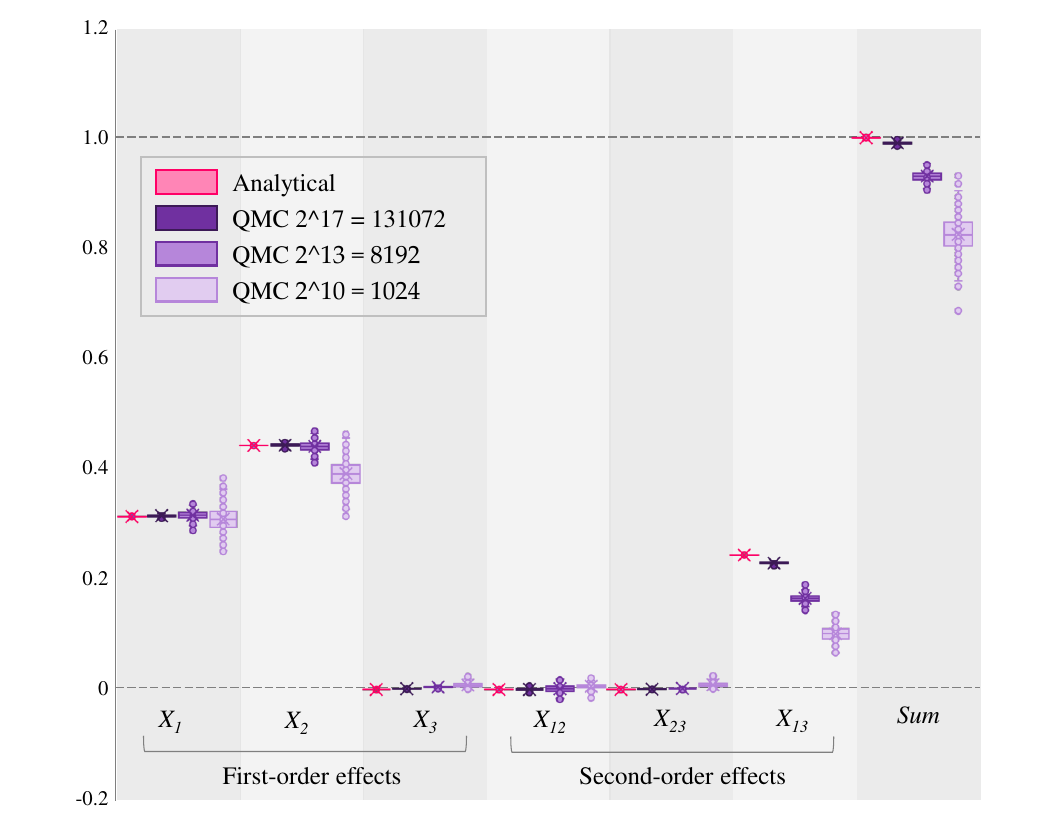}
\caption{\label{fig:ishigami_box}Analytical and approximate sensitivity indices for Ishigami function.}
\end{figure}

The simple binning algorithm is tested on the Ishigami function with parameters $a = 7$ and $b = 0.1$, quasi-random sampling of different sample size, Figure \ref{fig:ishigami_box}.

Figure \ref{fig:ishigami_box} indicates that the simple binning algorithm with QMC sample size $2^{13}$ or larger captures first-order effects very well, but underestimates the highly-curved (see Figure \ref{fig:ishigami_surf} (center)) interaction effect between $X_1$ and $X_3$. A smaller sample translates into fewer bins for estimating the first-order effects and the square root of that for second-order effects (a sample size of $2^{10}$ is analyzed with 10 bins for first-order effects and only 3 bins for the second-order effects), so the high-frequency relationships are downplayed. Thus, if a model is known to possess periodic or cyclic effects, a larger sample size becomes essential in order to produce reliable estimates for sensitivity analysis.

\subsection{Capturing correlation -- The mechanical engineering model}

In this section, we reproduce and then modify the mechanical engineering problem presented in \citet{marzban2016conceptual} by introducing dependent inputs into their model and examining the resulting performance capabilities of our sensitivity analysis algorithm in processing this additional complication. 

\subsubsection{Case background}

Civil engineering structures play an essential role in most modern infrastructures (e.g., in buildings, process industry, and shells). A wide spectrum of steel frames are required within these structures. From an engineering perspective, such frames must be designed to withstand numerous different kinds of load actions due to such things as payloads, self-weight, and environmental impacts. In statically loaded structures (characterized by permanent load actions), it is necessary to analyze stresses and displacements. Stress analyses are usually related to the ultimate limit state (ULS), while analyses on displacement cover the serviceability limit states (SLS) that are affiliated with the operational conditions and functionalities (service) of structures. It is crucial to understand the effects of different parameters on the mechanical system behavior. For some simple cases, the effects can be determined analytically. 

Frequently, however, nonlinear becomes impractical or impossible to solve analytically in even the most straightforward systems. In such cases, numerical modeling becomes a requisite tool for assessing structural behavior under mechanical loads. From an engineering viewpoint, it is usually easy to determine the most influential geometric and material parameters in the underlying linear analyses (LAs) of mechanical behavior of the system, such as deflections and stresses. However, geometrically nonlinear analyses (GNAs) may uncover different behavior of the system with a different set of influential parameters. Thus, the decision, which analysis to use is often critical for engineering design and structural analysis. \par

To support the decision as to whether a nonlinear approach is required, a simplified model that computes the ratio between the results of two different analysis types (e.g., LA and GNA) can be employed. The sensitivity indices for different parameters can then provide insight for decision-making. 

\begin{figure}[H]
\centering
\includegraphics[scale=1.0]{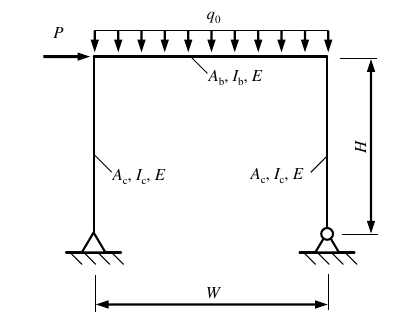}
\caption{\label{fig:frame}Shape and dimensions of the studied frame structure.}
\end{figure}

A two-dimensional frame structure is employed to demonstrate such an approach (Figure \ref{fig:frame}), as in (\citealp{marzban2016conceptual}). The frame structure is comprised of two vertical columns joined with a horizontal beam. The beam is loaded by a uniformly distributed shear load along the beam and the top left corner of the frame is loaded by a horizontal load. The structural behavior, focusing on the displacements, of the frame structure is numerically solved via LA and GNA.   \par

As the original CALFEM finite element (FE) code (\citealp{austrell2004calfem}) applied in \citet{marzban2016conceptual} was not accessible, a similar frame structure compiled by three beam elements was recreated (Figure \ref{fig:frame}). The numerical FE models were parameterized in an API tool using the FEMAP 2022.2 (Siemens PLM) software. The system was studied employing nine different input parameters. Namely, the height \(H\) and width  \(W\) of the frame; the modulus of elasticity \(E\); the cross-sectional area of columns and beam, \(A_c\) and \(A_b\), respectively; the second moment of areas of columns and beam, \(I_c\) and \(I_b\), respectively; lateral force, P; and uniformly distributed shear load \(q_0\). Each parameter was varied by $\pm 25 \%$ of its mean value and Table \ref{tab_eng_model} displays all of the corresponding minimum and maximum values. A total number of 10,000 models were simulated and analyzed using both LAs and GNAs. As an output value, the ratio between lateral displacements in the top left corner of the frame, obtained using GNA and LA, was used. \par

\begin{table}[H]
\centering 
\caption{\label{tab_eng_model}Model parameter ranges in the frame model.}
\begin{tabular}{llllllllll}
\hline
 &  \(H\)&  \(W\) &  \(E\)&  \(A_c\)&  \(I_c\)&  \(A_b\)&  \(I_b\)&  \(P\)&  \(q_0\) \\
  &  (m) & (m) & (GPa) & (e-3 m$^{2}$) & (e-5 m$^{4}$) & (e-3 m$^{2}$) & (e-5 m$^4$) & (kN) & (kN/m) \\
\hline
Minimum & 2.55 & 3.0 & 150 & 1.5 & 1.2 & 4.5 & 4.05 & 7.5 & 37.5 \\
Maximum & 4.25 & 5.0 & 250 & 2.5 & 2.0 & 7.5 & 6.75 & 12.5 & 62.5 \\                   
\hline
\end{tabular}
\end{table}

\subsubsection{Estimation of sensitivity indices}

Table \ref{tab_eng_1} displays the first- and second-order sensitivity indices computed for the original model used in \citet{marzban2016conceptual}. To clarify the visual perception, all index values less than 2\% have been grayed out. As can be seen from Table \ref{tab_eng_1}, the frame height is the mode influencing parameter. The relative magnitudes shown for the first-order effects coincide with the those obtained by \citet{marzban2016conceptual}. The negligible second-order indices signify an absence of any interaction effects.

\begin{table}[h]
\caption{\label{tab_eng_1} Sensitivity indices for the engineering model.}
\begin{tabular}{rrrrrrrrrr}
\hline
\multicolumn{1}{l}{}                     & \multicolumn{1}{l}{$H$}                       & \multicolumn{1}{l}{$W$}       & \multicolumn{1}{l}{$E$}       & \multicolumn{1}{l}{$A_c$}      & \multicolumn{1}{l}{$I_c$}      & \multicolumn{1}{l}{$A_b$}      & \multicolumn{1}{l}{$I_b$}      & \multicolumn{1}{l}{$P$}       & \multicolumn{1}{l}{$q_0$}      \\
\multicolumn{1}{l}{}                     & \multicolumn{1}{l}{(m)}                     & \multicolumn{1}{l}{(m)}     & \multicolumn{1}{l}{(N/$m^{2}$)}  & \multicolumn{1}{l}{(m$^{2}$)}    & \multicolumn{1}{l}{(m$^{4}$)}    & \multicolumn{1}{l}{(m$^{2}$)}    & \multicolumn{1}{l}{(m$^{4}$)}    & \multicolumn{1}{l}{(N)}     & \multicolumn{1}{l}{(N/m)}   \\ \hline
\multicolumn{1}{l}{First-order indices}  & 41 \%                                       & 16 \%                       & 13 \%                       & {\color[HTML]{9B9B9B} 0 \%} & 10 \%                       & {\color[HTML]{C0C0C0} 0 \%} & {\color[HTML]{C0C0C0} 0 \%} & {\color[HTML]{C0C0C0} 0 \%} & 13 \%                       \\ \hline
\multicolumn{1}{l}{}                     &                                             &                             &                             &                             &                             &                             &                             &                             &                             \\
\multicolumn{1}{l}{Second-order indices} &                                             &                             &                             &                             &                             &                             &                             &                             &                             \\ \hline
$H$ (m)                                    & {\color[HTML]{9B9B9B} }                     & {\color[HTML]{9B9B9B} 1 \%} & {\color[HTML]{9B9B9B} 1 \%} & {\color[HTML]{9B9B9B} 0 \%} & {\color[HTML]{9B9B9B} 0 \%} & {\color[HTML]{9B9B9B} 0 \%} & {\color[HTML]{9B9B9B} 0 \%} & {\color[HTML]{9B9B9B} 0 \%} & {\color[HTML]{9B9B9B} 1 \%} \\
$W$ (m)                                    & {\color[HTML]{9B9B9B} }                     & {\color[HTML]{9B9B9B} }     & {\color[HTML]{9B9B9B} 0 \%} & {\color[HTML]{9B9B9B} 0 \%} & {\color[HTML]{9B9B9B} 0 \%} & {\color[HTML]{9B9B9B} 0 \%} & {\color[HTML]{9B9B9B} 0 \%} & {\color[HTML]{9B9B9B} 0 \%} & {\color[HTML]{9B9B9B} 0 \%} \\
$E$ (N/m$^{2}$)                                 & {\color[HTML]{9B9B9B} }                     & {\color[HTML]{9B9B9B} }     & {\color[HTML]{9B9B9B} }     & {\color[HTML]{9B9B9B} 0 \%} & {\color[HTML]{9B9B9B} 0 \%} & {\color[HTML]{9B9B9B} 0 \%} & {\color[HTML]{9B9B9B} 0 \%} & {\color[HTML]{9B9B9B} 0 \%} & {\color[HTML]{9B9B9B} 0 \%} \\
$A_c$ (m$^{2}$)                                  & {\color[HTML]{9B9B9B} }                     & {\color[HTML]{9B9B9B} }     & {\color[HTML]{9B9B9B} }     & {\color[HTML]{9B9B9B} }     & {\color[HTML]{9B9B9B} 0 \%} & {\color[HTML]{9B9B9B} 0 \%} & {\color[HTML]{9B9B9B} 0 \%} & {\color[HTML]{9B9B9B} 0 \%} & {\color[HTML]{9B9B9B} 0 \%} \\
$I_c$ (m$^{4}$)                                  & {\color[HTML]{9B9B9B} }                     & {\color[HTML]{9B9B9B} }     & {\color[HTML]{9B9B9B} }     & {\color[HTML]{9B9B9B} }     & {\color[HTML]{9B9B9B} }     & {\color[HTML]{9B9B9B} 0 \%} & {\color[HTML]{9B9B9B} 0 \%} & {\color[HTML]{9B9B9B} 0 \%} & {\color[HTML]{9B9B9B} 0 \%} \\
$A_b$ (m$^{2}$)                                  & {\color[HTML]{9B9B9B} }                     & {\color[HTML]{9B9B9B} }     & {\color[HTML]{9B9B9B} }     & {\color[HTML]{9B9B9B} }     & {\color[HTML]{9B9B9B} }     & {\color[HTML]{9B9B9B} }     & {\color[HTML]{9B9B9B} 0 \%} & {\color[HTML]{9B9B9B} 0 \%} & {\color[HTML]{9B9B9B} 0 \%} \\
$I_b$ (m$^{4}$)                                  & {\color[HTML]{9B9B9B} }                     & {\color[HTML]{9B9B9B} }     & {\color[HTML]{9B9B9B} }     & {\color[HTML]{9B9B9B} }     & {\color[HTML]{9B9B9B} }     & {\color[HTML]{9B9B9B} }     & {\color[HTML]{9B9B9B} }     & {\color[HTML]{9B9B9B} 0 \%} & {\color[HTML]{9B9B9B} 0 \%} \\
$P$ (N)                                    & \multicolumn{1}{l}{{\color[HTML]{9B9B9B} }} & {\color[HTML]{9B9B9B} }     & {\color[HTML]{9B9B9B} }     & {\color[HTML]{9B9B9B} }     & {\color[HTML]{9B9B9B} }     & {\color[HTML]{9B9B9B} }     & {\color[HTML]{9B9B9B} }     & {\color[HTML]{9B9B9B} }     & {\color[HTML]{9B9B9B} 0 \%} \\ \hline
\end{tabular}
\end{table}

To introduce dependency into the system, the \(H\)/\(I_c\) ratio was fixed as per the dimensions of the case structure (i.e., \(I_c\) = 4.706e-6$\cdot$\(H\)). With this change in effect, to investigate the second-order effects of input parameters with correlated input parameters, a subsequent computational experiment was conducted with a total of 10,000 simulations. The new sensitivity indices calculated for this correlated system are presented in Table \ref{tab_eng_2}. \par

\begin{table}[h]
\caption{\label{tab_eng_2} Sensitivity indices for the modified engineering model with correlated inputs.}
\begin{tabular}{rrrrrrrrrr}
\hline
\multicolumn{1}{l}{}                     & \multicolumn{1}{l}{$H$}                       & \multicolumn{1}{l}{$W$}                       & \multicolumn{1}{l}{$E$}                       & \multicolumn{1}{l}{$A_c$}                      & \multicolumn{1}{l}{$I_c$}                      & \multicolumn{1}{l}{$A_b$}                      & \multicolumn{1}{l}{$I_b$}                      & \multicolumn{1}{l}{$P$}                       & \multicolumn{1}{l}{$q_0$}      \\
\multicolumn{1}{l}{}                     & \multicolumn{1}{l}{(m)}                     & \multicolumn{1}{l}{(m)}                     & \multicolumn{1}{l}{(N/m$^{2}$)}                  & \multicolumn{1}{l}{(m$^{2}$)}                    & \multicolumn{1}{l}{(m$^{4}$)}                    & \multicolumn{1}{l}{(m$^{2}$)}                    & \multicolumn{1}{l}{(m$^{4}$)}                    & \multicolumn{1}{l}{(N)}                     & \multicolumn{1}{l}{(N/m)}   \\ \hline
\multicolumn{1}{l}{First-order indices}  & 21 \%                                       & 27 \%                                       & 23 \%                                       & {\color[HTML]{9B9B9B} 0 \%}                 & 21 \%                                       & {\color[HTML]{C0C0C0} 0 \%}                 & {\color[HTML]{C0C0C0} 1 \%}                 & {\color[HTML]{C0C0C0} 0 \%}                 & 22 \%                       \\ \hline
\multicolumn{1}{l}{}                     &                                             &                                             &                                             &                                             &                                             &                                             &                                             &                                             &                             \\
\multicolumn{1}{l}{Second-order indices} &                                             &                                             &                                             &                                             &                                             &                                             &                                             &                                             &                             \\ \hline
$H$ (m)                                    & {\color[HTML]{9B9B9B} }                     & {\color[HTML]{9B9B9B} 1 \%}                 & {\color[HTML]{9B9B9B} 0 \%}                 & {\color[HTML]{9B9B9B} 0 \%}                 & {\color[HTML]{000000} -20 \%}               & {\color[HTML]{9B9B9B} 0 \%}                 & {\color[HTML]{9B9B9B} 0 \%}                 & {\color[HTML]{9B9B9B} 0 \%}                 & {\color[HTML]{9B9B9B} 2 \%} \\
$W$ (m)                                    & {\color[HTML]{9B9B9B} }                     & {\color[HTML]{9B9B9B} }                     & {\color[HTML]{9B9B9B} 0 \%}                 & {\color[HTML]{9B9B9B} 0 \%}                 & {\color[HTML]{9B9B9B} 1 \%}                 & {\color[HTML]{9B9B9B} 0 \%}                 & {\color[HTML]{9B9B9B} 0 \%}                 & {\color[HTML]{9B9B9B} 0 \%}                 & {\color[HTML]{9B9B9B} 0 \%} \\
$E$ (N/m$^{2}$)                                 & {\color[HTML]{9B9B9B} }                     & {\color[HTML]{9B9B9B} }                     & {\color[HTML]{9B9B9B} }                     & {\color[HTML]{9B9B9B} 0 \%}                 & {\color[HTML]{9B9B9B} 0 \%}                 & {\color[HTML]{9B9B9B} 0 \%}                 & {\color[HTML]{9B9B9B} 0 \%}                 & {\color[HTML]{9B9B9B} 0 \%}                 & {\color[HTML]{9B9B9B} 0 \%} \\
$A_c$ (m$^{2}$)                                  & {\color[HTML]{9B9B9B} }                     & {\color[HTML]{9B9B9B} }                     & {\color[HTML]{9B9B9B} }                     & {\color[HTML]{9B9B9B} }                     & {\color[HTML]{9B9B9B} 0 \%}                 & {\color[HTML]{9B9B9B} 0 \%}                 & {\color[HTML]{9B9B9B} 0 \%}                 & {\color[HTML]{9B9B9B} 0 \%}                 & {\color[HTML]{9B9B9B} 0 \%} \\
$I_c$ (m$^{4}$)                                  & {\color[HTML]{9B9B9B} }                     & {\color[HTML]{9B9B9B} }                     & {\color[HTML]{9B9B9B} }                     & {\color[HTML]{9B9B9B} }                     & {\color[HTML]{9B9B9B} }                     & {\color[HTML]{9B9B9B} 0 \%}                 & {\color[HTML]{9B9B9B} 0 \%}                 & {\color[HTML]{9B9B9B} 0 \%}                 & {\color[HTML]{9B9B9B} 2 \%} \\
$A_b$ (m$^{2}$)                                  & {\color[HTML]{9B9B9B} }                     & {\color[HTML]{9B9B9B} }                     & {\color[HTML]{9B9B9B} }                     & {\color[HTML]{9B9B9B} }                     & {\color[HTML]{9B9B9B} }                     & {\color[HTML]{9B9B9B} }                     & {\color[HTML]{9B9B9B} 0 \%}                 & {\color[HTML]{9B9B9B} 0 \%}                 & {\color[HTML]{9B9B9B} 0 \%} \\
$I_b$ (m$^{4}$)                                  & {\color[HTML]{9B9B9B} }                     & {\color[HTML]{9B9B9B} }                     & {\color[HTML]{9B9B9B} }                     & {\color[HTML]{9B9B9B} }                     & {\color[HTML]{9B9B9B} }                     & {\color[HTML]{9B9B9B} }                     & {\color[HTML]{9B9B9B} }                     & {\color[HTML]{9B9B9B} 0 \%}                 & {\color[HTML]{9B9B9B} 0 \%} \\
$P$ (N)                                    & \multicolumn{1}{l}{{\color[HTML]{9B9B9B} }} & \multicolumn{1}{l}{{\color[HTML]{9B9B9B} }} & \multicolumn{1}{l}{{\color[HTML]{9B9B9B} }} & \multicolumn{1}{l}{{\color[HTML]{9B9B9B} }} & \multicolumn{1}{l}{{\color[HTML]{9B9B9B} }} & \multicolumn{1}{l}{{\color[HTML]{9B9B9B} }} & \multicolumn{1}{l}{{\color[HTML]{9B9B9B} }} & \multicolumn{1}{l}{{\color[HTML]{9B9B9B} }} & {\color[HTML]{9B9B9B} 0 \%} \\ \hline
\end{tabular}
\end{table}

Comparing Tables \ref{tab_eng_1} and \ref{tab_eng_2}, we can clearly observe the striking difference arising in the second-order effects. While for the original model, all second-order effects were negligible, in the correlated model, we can see a negative effect of 20\%. The negative second-order effect reflects the introduced dependency between the inputs. It should be noted that this negative second-order effect essentially zeroes out the first-order effect of the $I_c$ if summing up all effects. Thus, the negative second-order effect represents a correction of an overlapping first-order effect.  

It can be further observed that the first-order indices have changed their values as well. By effectively eliminating one input variable via correlation, the overall output variance of the output has changed so that corresponding portions of the explained variance have also changed accordingly.  

This engineering case demonstrates that the binning approach for calculating sensitivity indices is capable of capturing and identifying the impacts from dependent inputs. 

Even though such sensitivity analyses can be time-consuming for complex systems (e.g., for a high number of elements), this approach could be used for sub-models to highlight the influencing factors and importance of GNAs. While in the presented example, the correlation was introduced artificially, it occurs naturally in more complex systems (e.g., strength properties are dependent on the material chosen, or, if giving an example outside engineering applications, different price ranges lead to different demand levels, etc.). Projecting these findings to a larger scope of applications,an increase in the accuracy of structural analyses for SLS estimates would improve the overall reliability and integrity of the structures. Consequently, costly required changes to structural elements potentially found \textit{post hoc} during service/operation could be avoided via proper \textit{a priori} engineering analysis and design.

\section{Relationships between second-order indices and correlation}
\label{s:corr}

The mechanical engineering model demonstrated that in the presence of positive correlation, the second-order sensitivity index turns negative (Table \ref{tab_eng_2}). But how does the index behave when the correlation is negative? Furthermore, is the relationship the same in the presence of an interaction of inputs that affect the output?

To explore these questions, we set up a series of simulation experiments for two simple two-factor models. The first model is additive, $Y=A+B$, where the variables do not interact. The second model is multiplicative, $Y=AB$, which possesses a positive second-order effect when the variables are independent, thereby manifesting interaction. We also examine two different types of correlation, (i) joint multivariate uniform distribution copula and (ii) equating values of $B$ to $A$ over a portion of the sample, which might be considered as a structural change in the system. Both correlation types are modeled for different correlation strengths. For validation purposes, Pearson and Spearman correlation coefficients are reported in each case. Both $A$ and $B$ assume uniform distribution between $0$ and $5$. The indices are calculated based on a sample size of $10^6$ to ensure a high level of precision. 

\begin{figure}[H]
\centering
\includegraphics[scale=0.6]{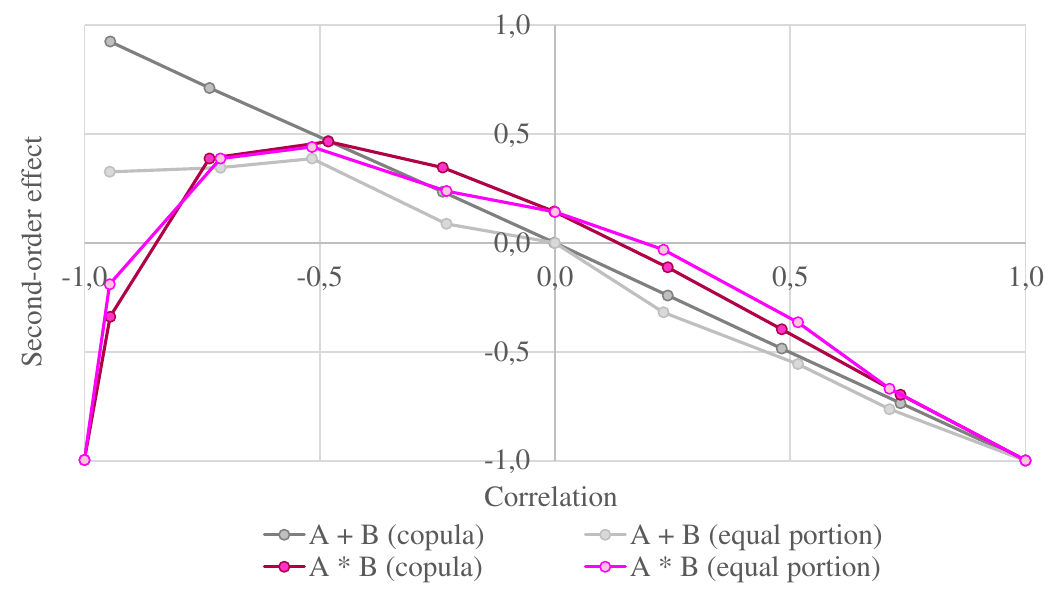}
\caption{\label{fig:corr}Second-order effect as a function of correlation for additive (0\% interaction) and multiplicative (14\% interaction) functions with two correlation types, 
 (1) - portion of equal values, (2) - copula.}
\end{figure}

The sensitivity indices estimates are presented in Table \ref{table:corr}, and the relationship between the second-order indices and the Pearson correlation coefficient is visualized in Figure \ref{fig:corr}. \citet{hart2018approximation} have indicated that theoretical second-order indices can become negative in the presence of dependent inputs and such an outcome can be observed for the approximated indices computed by our simple binning algorithm (Table \ref{table:corr}).

\begin{table}[h]
\caption{\label{table:corr} Sensitivity indices and correlation coefficients for a simple $Y=AB$ model with $A$ and $B$ dependent to different degrees.}

\begin{tabular}{lllrrrrrrrrr}
\hline
\multirow{3}{*}{\begin{tabular}[c]{@{}l@{}}Correlation\\ type\end{tabular}} & \multirow{3}{*}{Model} & \multirow{3}{*}{$S$} & \multicolumn{9}{c}{Correlation}                                                                                                                                                                                                                                          \\ \cline{4-12} 
                                  &                        &                          & \multicolumn{4}{c}{negative}                                                                                                          & \multicolumn{1}{c}{}     & \multicolumn{4}{c}{positive}                                                                           \\
                                  &                        &                          & \multicolumn{1}{l}{-100 \%} & \multicolumn{1}{l}{…} & \multicolumn{1}{l}{-50 \%} & \multicolumn{1}{l}{…} & \multicolumn{1}{l}{0 \%} & \multicolumn{1}{l}{…} & \multicolumn{1}{l}{50 \%} & \multicolumn{1}{l}{…} & \multicolumn{1}{l}{100 \%} \\ \hline
\multirow{10}{*}
{\includegraphics[width=0.85in]{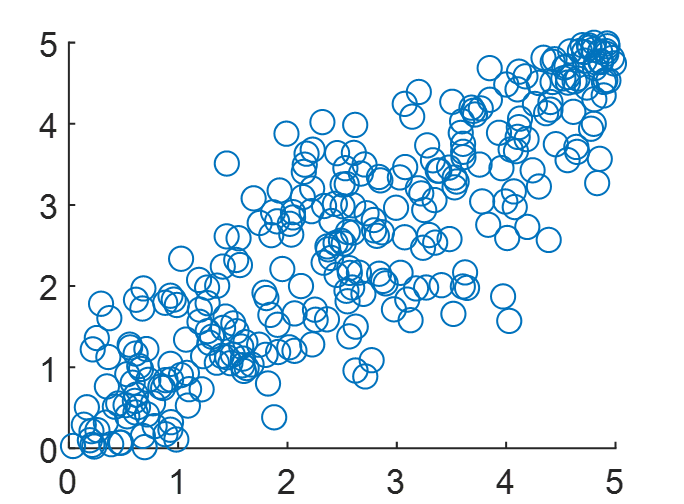}} & \multirow{4}{*}{$A+B$} & $S_A$                     & \multicolumn{1}{r}{-}         & 0.14                  & 0.27                       & 0.38                  & 0.50                     & 0.62                  & 0.74                      & 0.87                  & 1.00                       \\
                                  &                        & $S_B$                     & \multicolumn{1}{r}{-}               & 0.14                  & 0.27                       & 0.38                  & 0.50                     & 0.62                  & 0.74                      & 0.87                  & 1.00                       \\
                                  &                        & $S_{AB}$                    & \multicolumn{1}{r}{-}               & \textbf{0.71}         & \textbf{0.47}              & \textbf{0.23}         & \textbf{0.00}            & \textbf{-0.24}        & \textbf{-0.49}            & \textbf{-0.74}        & \textbf{-1.00}             \\
                                  &                        & $\sum$                       & \multicolumn{1}{r}{-}                       & 1.00                  & 1.00                       & 1.00                  & 1.00                     & 1.00                  & 1.00                      & 1.00                  & 1.00                       \\ \cline{2-12} 
                                  & \multirow{4}{*}{$AB$} & $S_A$                     & 1.00                                       & 0.31                  & 0.27                       & 0.33                  & 0.43                     & 0.56                  & 0.70                      & 0.85                  & 1.00                       \\
                                  &                        & $S_B$                     & 1.00                                           & 0.31                  & 0.27                       & 0.33                  & 0.43                     & 0.56                  & 0.70                      & 0.85                  & 1.00                       \\
                                  &                        & $S_{AB}$                    & \textbf{-1.00}                         & \textbf{0.39}         & \textbf{0.47}              & \textbf{0.35}         & \textbf{0.14}            & \textbf{-0.11}        & \textbf{-0.40}            & \textbf{-0.70}        & \textbf{-1.00}             \\
                                  &                        & $\sum$                      & 1.00                                      & 1.00                  & 1.00                       & 1.00                  & 1.00                     & 1.00                  & 1.00                      & 1.00                  & 1.00                       \\ \cline{2-12} 
                                  & \multicolumn{2}{l}{Pearson}           & -1.00                                     & -0.73                 & -0.48                      & -0.24                 & 0.00                     & 0.24                  & 0.48                      & 0.73                  & 1.00                       \\
                                  & \multicolumn{2}{l}{Spearman}          & -1.00                                  & -0.73                 & -0.48                      & -0.24                 & 0.00                     & 0.24                  & 0.48                      & 0.73                  & 1.00                       \\ \hline
\multirow{9}{*}{\includegraphics[width=0.85in]{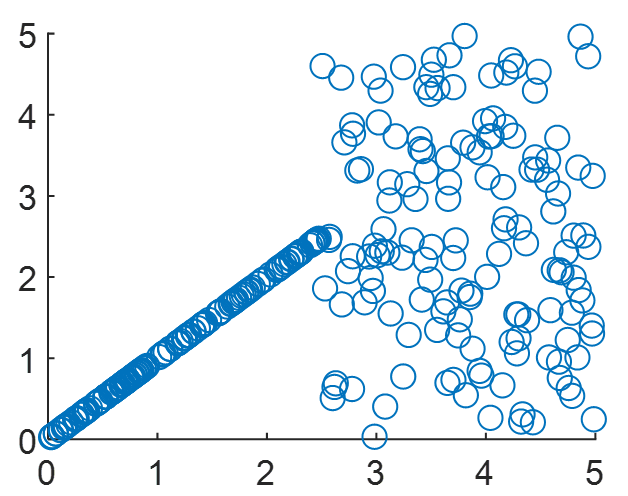}}   & \multirow{4}{*}{$A+B$} & $S_A$                     &   -                                        & 0.61                  & 0.36                       & 0.45                  & 0.50                     & 0.65                  & 0.79                      & 0.93                  & 1.00                       \\
                                  &                       & $S_B$                     &  -                                     & 0.04                  & 0.26                       & 0.47                  & 0.50                     & 0.67                  & 0.76                      & 0.83                  & 1.00                       \\
                                  &                        & $S_{AB}$                    &      -                            & \textbf{0.35}         & \textbf{0.39}              & \textbf{0.09}         & \textbf{0.00}            & \textbf{-0.32}        & \textbf{-0.56}            & \textbf{-0.76}        & \textbf{-1.00}             \\
                                  &                       & $\sum$                       &   -                                  & 1.00                  & 1.00                       & 1.00                  & 1.00                     & 1.00                  & 1.00                      & 1.00                  & 1.00                       \\ \cline{2-12} 
                                  & \multirow{4}{*}{$AB$} & $S_A$                     & 1.00                                             & 0.52                  & 0.39                       & 0.42                  & 0.43                     & 0.44                  & 0.51                      & 0.74                  & 1.00                       \\
                                  &                        & $S_B$                     & 1.00                                    & 0.09                  & 0.17                       & 0.35                  & 0.43                     & 0.59                  & 0.86                      & 0.93                  & 1.00                       \\
                                  &                        & $S_{AB}$                    & \textbf{-1.00}                        & \textbf{0.39}         & \textbf{0.44}              & \textbf{0.24}         & \textbf{0.14}            & \textbf{-0.03}        & \textbf{-0.36}            & \textbf{-0.67}        & \textbf{-1.00}             \\
                                  &                        & $\sum$                       & 1.00                                  & 1.00                  & 1.00                       & 1.01                  & 1.00                     & 1.00                  & 1.00                      & 1.00                  & 1.00                       \\ \cline{2-12} 
                                  & \multicolumn{2}{l}{Pearson}           & -1.00                                      & -0.71                 & -0.52                      & -0.23                 & 0.00                     & 0.23                  & 0.52                      & 0.71                  & 1.00                       \\
                                  & \multicolumn{2}{l}{Spearman}          & -1,00                               & -0.71                 & -0.53                      & -0.24                 & 0.00                     & 0.24                  & 0.53                      & 0.71                  & 1.00                       \\ \hline
\end{tabular}
\end{table}

Additive models cross the origin (see Figure \ref{fig:corr}) and possess zero second-order effects under a no-dependency case. No interaction is anticipated from the additive model. Copula variables show the second-order effect values very close to the correlation coefficients and form a linear relationship between the two. Positive correlation results in negative second-order effects, conveying overlapping effects of the variables. Conversely, negative correlation results in a positive second-order effect for additive models as if the combined effect of the variables is greater than the sum of their individual effects (see equation \ref{eq:AB}). The additive model, in which dependence is modeled through a portion of equal values, has largely the same pattern except for lower positive second-order effects in the presence of negative correlation. The calculation of significance indices for both additive models at -100\% correlation fails, because all $Y$ values turn to $0$. Instead, the values for the negative 95\% correlation are computed and displayed in Figure \ref{fig:corr}.         

In the multiplicative model, the second-order effect index behaves asymmetrically for positive and negative correlation (Figure \ref{fig:corr}). With increasing positive correlation, the second-order index becomes increasingly negative (as in the additive models), signifying the overlapping effects of the input variables on the output. For negative correlations, the second-order index is initially positive and increasing, peaking at 50\% correlation. This translates into a synergistic effect where the negatively correlated inputs magnify the impact on the output. But when the negative correlation exceeds 50\%, the second-order index decreases and approaches $-1$. One could speculate that with a strong negative correlation for interacting variables, the joint overlapping effect becomes stronger than the synergistic one, driving the second-order index to $-1$. 

For all cases, the \textit{conservation} property (the sum of all indices is equal to 1) holds. The \textit{boundedness} property for each index extends from $[0, 1]$, as for the classic Sobol' indices, to $[-1, 1]$. Because the conservation property holds under all cases, this implies that correlation and interaction effects are actually both combined into the single second-order effect, and, simultaneously affecting the first-order index behavior, perfectly fitting into the variance decomposition of the output. The presence of interaction can be observed in Figure \ref{fig:corr} as the vertical distance between the additive and multiplicative models' lines on the correlation range between $-0.25$ and $0.25$. In other words, the different degrees of dependency move the second-order effect up or down, but the presence of interaction reveals itself in higher second-order effect values.

\section{Visualizing second-order effects -- Beyond sensitivity indices}
\label{s:simdec}

The sensitivity analysis field has focused on computing sensitivity indices to be able to rank the input variables by their relative importance (\citealp{saltelli2008global,da2021basics}). And yet it becomes apparent that sensitivity indices are not able to convey the full relationship story. As observed in our earlier analysis with dependent input variables (Table \ref{table:corr}), the second-order indices for the same model, but operating under different assumptions, appear to be very similar. However,  the scatter plots (Table \ref{table:corr}) reveal how different these relationships actually are, and this knowledge can lead to different structural design decisions. 

A fact that has been largely neglected throughout modern sensitivity analysis studies is that although specific summary values can appear near-identical in magnitude, an examination of the underlying data can reveal very different shapes and perspectives (\citealp{puy2022models, saltelli2004sensitivity, kozlova2024uncovering}).

Consequently, different kinds of visualizations can be used to augment and improve the understanding of the underlying model. These data visualizations can include such formats as basic scatter plots, response surfaces, heat maps, and parallel coordinate plots. However, all of these representations possess dimensionality limitations. SimDec is a recent approach that is able to project multidimensional systems onto a two-dimensional graph (\citealp{kozlova2022monte}) and, in combination with the simple binning sensitivity analysis approach, can be used to convey the nature of interactions in complex engineering models to produce a much more comprehensive sensitivity analysis framework (\citealp{kozlova2024uncovering}).

To illustrate the power of this framework, this section showcases the application of the simple binning approach together with SimDec visualization for investigating the behavior of a complex structural reliability model containing both correlations and interactions.

\subsection{Case background: structural reliability model for fatigue assessments}

Many structural applications experience cyclic or fluctuating load conditions during service life (such as ship and offshore structures as well as various vehicles, trucks and trains). In these complex systems, fatigue is amongst the most important design criteria to control and maintain the lifespan of ageing components without concerns of structural failures. Due to the random nature of acting cyclic loads in many applications and the complexity of physical modeling of fatigue phenomenon, the vast majority of existing models for predicting fatigue life are necessarily based on probabilistic or data-driven approaches (\citealp{bai2023ress, Ye2023ress,  ruiz2020ress,leonetti2020ijf}). However, current trends for improving material usage and optimization in mechanical components have led to an ever-increasing need for creating more sophisticated and more accurate deterministic models to assess the fatigue life of components. This is particularly the case for modern infrastructures that utilize high-strength steels, due to the fact that an increase in material strength does not contribute to higher fatigue performance in welded components (\citealp{lieurade2008WitW}). 

To account for the effects of material strength and welding quality, a multi-parametric model -- named the 4R method -- was introduced for fatigue assessment of welded components. The 4R model incorporates four parameters - material strength, welding residual stresses, applied stress ratio, and welding quality (\citealp{ahola2020ijf, ahola2020jcsr-4R}). These four key parameters affect the fatigue strength of welded components depending on the conditions (\citealp{hobbacher2016springer}). As the 4R model employs more parameters than conventional models, a key question still rests on a determination of the sensitivity of each parameter on the 4R output value.

\begin{figure}[H]
\centering
\includegraphics[scale=1.0]{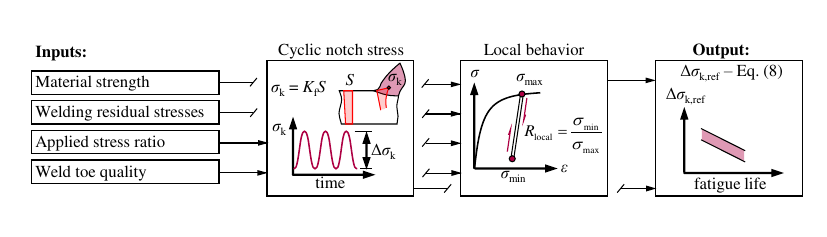}
\caption{\label{fig:4R}Description and workflow of the 4R model for fatigue assessment.}
\end{figure}

The basis of the 4R method is to utilize the effective stress concept, usually known as the effective notch stress (ENS) approach (\citealp{sonsino2012ijf}), in association with the mean stress correction using the requisite four parameters. In the 4R model, material elastic-plastic behavior at the notch root is analytically or numerically computed, after which the local hysteresis loops of cyclic stress are determined. The output reference stress applied in fatigue assessments (in the form of S-N curve) is computed based on the well-known Smith-Watson-Topper (\citealp{swt1970}) mean stress correction on the linear-elastic notch stress (equation \ref{eq:4Rrefstress}). Figure \ref{fig:4R} demonstrates the workflow of the 4R method.

\begin{equation} \label{eq:4Rrefstress}
\ \Delta\sigma_{k,ref} = \frac{\Delta\sigma_{k}}{\sqrt{1-R_{local}}} \\
\end{equation}

\subsection{Sensitivity indices for the structural reliability model}

The 4R model is simulated $10^4$ times with inputs uniformly distributed within the ranges specified in Table \ref{tab_4R_inputs}.

\begin{table}[h]
\caption{\label{tab_4R_inputs} Assumptions on input variation for the structural reliability 4R model.}
\begin{tabular}{lllll}
\hline
Case material         & \begin{tabular}[c]{@{}l@{}}Weld toe radius,\\ $r_{true}$ (mm)\end{tabular} & \begin{tabular}[c]{@{}l@{}}Steel grade,\\$R_{p0.2}$ (MPa)\end{tabular} & \begin{tabular}[c]{@{}l@{}}Residual stress, \\$\sigma_{res}$ (MPa)\end{tabular} & \begin{tabular}[c]{@{}l@{}}Stress ratio,\\$R$ (-)\end{tabular} \\ \hline
\multirow{3}{*}{S355} & \multirow{3}{*}{0.5 ± 0.5}                                        & \multirow{3}{*}{355 ± 100}                                     & 250 ± 100                                                           & 0.5 ± 0.2                                                      \\
                      &                                                                   &                                                                & 0 ± 100                                                             & 0 ± 0.2                                                        \\
                      &                                                                   &                                                                & -100 ± 100                                                          & -1 ± 0.2                                                       \\ \hline
\multirow{3}{*}{S960} & \multirow{3}{*}{0.5 ± 0.5}                                        & \multirow{3}{*}{960 ± 100}                                     & 850 ±   100                                                         & 0.5 ±   0.2                                                    \\
                      &                                                                   &                                                                & 0 ± 100                                                             & 0 ± 0.2                                                        \\
                      &                                                                   &                                                                & -300 ± 100                                                          & -1 ± 0.2                                                       \\ \hline
\end{tabular}
\end{table}

The three levels of residual stress and stress ratio for each case material are formed based on actual conditions in structural components. Usually welding causes high tensile residual stresses, up to the yield strength of parent material. On the other hand, residual stresses can be reduced by, for example, post-heat treatments and/or mechanical post-weld treatments or residual stress relaxation via the mechanical load. The applied stress ratio is specific to load conditions in service, and the selected three cases represent available cases in engineering components. The weld toe radius is converted in the model to fatigue-effective stress, $K_f$, the sensitivity to which is analyzed. The three other inputs used in the sensitivity analysis, as is. The sensitivity indices to the model output, $\Delta\sigma_{k,ref}$, computed with the simple binning method, are presented in Table \ref{tab_4R_SI}.

\begin{table}[h]
\caption{\label{tab_4R_SI} Sensitivity indices for the structural reliability 4R model.}
\begin{tabular}{lrrrrrr}
\hline
\multirow{2}{*}{Variable}    & \multicolumn{1}{l}{\multirow{2}{*}{\begin{tabular}[c]{@{}l@{}}First-order\\ effect\end{tabular}}} & \multicolumn{4}{c}{Second-order effect}                                                                                       & \multicolumn{1}{l}{\multirow{2}{*}{\begin{tabular}[c]{@{}l@{}}Combined\\ effect\end{tabular}}} \\ \cline{3-6}
                             & \multicolumn{1}{l}{}                                                                              & \multicolumn{1}{c}{\begin{tabular}[c]{@{}c@{}}Residual\\ stress,\\ $\sigma_{res}$\end{tabular}} & \multicolumn{1}{c}{\begin{tabular}[c]{@{}c@{}}Stress\\ ratio,\\ $R$\end{tabular}} & \multicolumn{1}{c}{\begin{tabular}[c]{@{}c@{}}Steel\\ grade,\\ $R_{p0.2}$\end{tabular}} & \multicolumn{1}{c}{\begin{tabular}[c]{@{}c@{}}Fatigue-\\ effective\\ stress, $K_f$\end{tabular}} & \multicolumn{1}{l}{}                                                                           \\ \hline
Residual   stress, $\sigma_{res}$    & 50\%                                                                                             &                                                                                         & 11\%                                                                           & $-6\%$                                                                              & 0\%                                                                                          & 51\%                                                                                          \\
Stress   ratio, $R$            & 28\%                                                                                             &                                                                                         &                                                                                 & 4\%                                                                               & 0\%                                                                                          & 35\%                                                                                          \\
Steel grade, $R_{p0.2}$           & 11\%                                                                                             &                                                                                         &                                                                                 &                                                                                    & 0\%                                                                                          & 10\%                                                                                          \\
Fatigue-effective stress, $K_f$ & 4\%                                                                                              &                                                                                         &                                                                                 &                                                                                    &                                                                                               & 4\%                                                                                           \\
$\Sigma$                       & 92\%                                                                                             &                                                                                         &                                                                                 &                                                                                    &                                                                                               & 100\%                                                                                         \\ \hline
\end{tabular}
\end{table}

The sensitivity indices in Table \ref{tab_4R_SI} show that the residual stress $\sigma_{res}$, alone, explains half of the variance of the output. Residual stress $\sigma_{res}$ also has $11\%$ interaction with stress ratio $R$, and a negative $-6\%$ second-order effect with steel grade $R_{p0.2}$. In addition, stress ratio $R$ and steel grade $R_{p0.2}$ have a $4\%$ interaction. The remaining second-order effects are zero. Thus, there are three input variables in the model that have pair-wise interaction effects with each other: residual stress $\sigma_{res}$, stress ratio $R$, and steel grade $R_{p0.2}$. Together these variables explain $96\%$ of the output variability. However, as mentioned above, the sensitivity indices alone cannot adequately explain the nature of those interactions and how they should be accounted for in decision-making processes.

\subsection{Exploring the nature of second-order effects}

To investigate how the model output is affected simultaneously by the three input variables identified as possessing pairwise interaction effects, we employ SimDec to provide a multidimensional visualization perspective of their impacts (\citealp{kozlova2022extending,kozlova2024uncovering}). Since SimDec uses a frequency distribution, the same simulation data that was used to compute sensitivity indices with the simple binning method, can be used for the visualization. The idea behind SimDec is to decompose the simulation data into regions formed by the combinations of different ranges (or states) of influential input variables. An intelligent color-coding of these regions in the frequency distribution communicates a clear visual representation of  the inherent input-output relationships within the model (for the algorithm details, see (\citealp{kozlova2024uncovering}). The open-source SimDec code is available for Python, R, Julia, and Matlab (\citealp{simdecomp2023}).

For the structural reliability model, we choose to break down the residual stress $\sigma_{res}$ into three states (low, medium, high), the stress ratio $R$ into two states (reversed, pulsating), and the steel grade $R_{p0.2}$ into two states (mild, UHSS) (Table \ref{tab_4R_states}).

\begin{table}[h]
\caption{\label{tab_4R_states} Decomposition set-up for the structural reliability 4R model (UHSS stands for ultra-high steel strength).}
\begin{tabular}{lllllllllll}
\hline
\multicolumn{3}{l}{Residual stress}    &  & \multicolumn{3}{l}{Stress ratio} &  & \multicolumn{3}{l}{Steel grade} \\ \cline{1-3} \cline{5-7} \cline{9-11} 
State                     & min  & max &  & State         & min     & max    &  & State      & min     & max      \\ \hline
Low & $-400$ & 100 &  & Reversed      & $-1.2$    & $-0.25$  &  & Mild       & 255     & 657      \\
Medium                    & 100  & 650 &  & Pulsating   & $-0.25$   & 0.7    &  & UHSS       & 657     & 1060     \\
High                      & 650  & 950 &  & -             & -       & -      &  & -          & -       & -        \\ \hline
\end{tabular}
\end{table}

 Altogether, all possible combinations of these state settings for the input variables (Table \ref{tab_4R_states}) form 12 separate scenarios.  The probability distribution of the model output, $\Delta\sigma_{k,ref}$, is partitioned using these scenarios and color-coded. The coloring logic is important to facilitate the overall interpretability of the visual perceptions. The states of the most influential input variable assume distinct main colors, and each of these main colors is sub-shaded to further highlight the partitions, Figure \ref{fig:simdec}.

\begin{table}[htpb!]
    \begin{tabularx}{\linewidth}{*{1}{>{\centering\arraybackslash}X}}
        \begin{center}
            \includegraphics[width=0.8\textwidth]{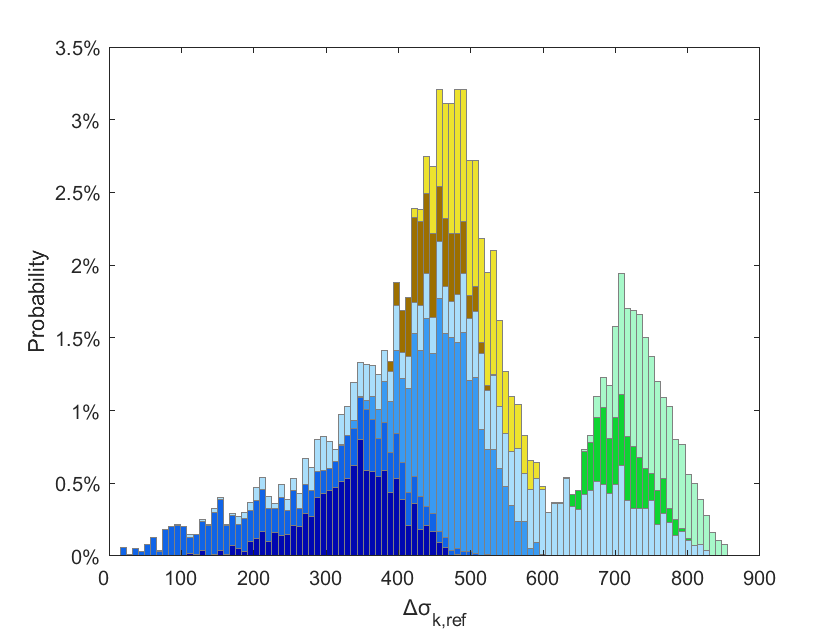}
            \captionof{figure}{\textit{Simulation decomposition} of the structural reliability 4R model.}
            \label{fig:simdec}
        \end{center}
        \\
        \centering
        \scriptsize
        \begin{tabular}{llcccrrrr}
            \toprule
            & & & & & \multicolumn{4}{c}{\textbf{Output Summary}} \\
            \cmidrule{6-9}
            \multirow{-2}{*}{\textbf{Color}} & \multirow{-2}{*}{\textbf{Scenario}} & \multirow{-2}{*}{\makecell{\begin{tabular}{l}\bf Residual\\stress, $\sigma_{res}$ \end{tabular}}} & \multirow{-2}{*}{\makecell{\begin{tabular}{l}\bf Stress\\ratio, $R$ \end{tabular}}}  & \multirow{-2}{*}{\makecell{\begin{tabular}{l}\bf Steel\\grade, $R_{p0.2}$ \end{tabular}}}  & min           & mean        & max        & probability     \\
            \midrule
            \cellcolor[HTML]{0008B1}                        & sc1                              &                                 &                        & Mild                                  & $107$        & $332$      & $475$     & $11\%$         \\
            \cellcolor[HTML]{1064E7}                        & sc2                              &                                 & \multirow{-2}{*}{Reversed}  & UHSS                                 & $11$        & $264$      & $515$     & $11\%$         \\
            \cellcolor[HTML]{399AF2}                        & sc3                              &                                 &                        & Mild                                  & $320$        & $467$      & $591$     & $22\%$         \\
            \cellcolor[HTML]{A9DEFB}                        & sc4                              & \multirow{-4}{*}{\begin{tabular}[c]{@{}l@{}}Low \end{tabular}}         & \multirow{-2}{*}{Pulsating} & UHSS                                 & $68$        & $543$      & $819$     & $22\%$         \\
            \midrule
            \cellcolor[HTML]{9C6F00}                        & sc5                              &                                 &                        & Mild                                  & $383$        & $456$      & $524$     & $6\%$         \\
            \cellcolor[HTML]{DFCB0B}                        & \textcolor{lightgray}{sc6}       &           & \multirow{-2}{*}{Reversed}  & \textcolor{lightgray}{UHSS}       & \textcolor{lightgray}{NaN}        & \textcolor{lightgray}{NaN}      & \textcolor{lightgray}{NaN}     & \textcolor{lightgray}{NaN}         \\
            \cellcolor[HTML]{EDE22E}                        & sc7                              &                                 &                        & Mild                                  & $422$        & $499$      & $622$     & $12\%$         \\
            \cellcolor[HTML]{FAF7A0}                        & \textcolor{lightgray}{sc8}       & \multirow{-4}{*}{Medium}           & \multirow{-2}{*}{Pulsating} & \textcolor{lightgray}{UHSS}                                 & \textcolor{lightgray}{NaN}        & \textcolor{lightgray}{NaN}      & \textcolor{lightgray}{NaN}     & \textcolor{lightgray}{NaN}         \\
            \midrule
            \cellcolor[HTML]{008201}                        & \textcolor{lightgray}{sc9}       &              &          & \textcolor{lightgray}{Mild}                 & \textcolor{lightgray}{NaN}        & \textcolor{lightgray}{NaN}      & \textcolor{lightgray}{NaN}     & \textcolor{lightgray}{NaN}         \\
            \cellcolor[HTML]{0ED432}                        & sc10                             &              & \multirow{-2}{*}{Reversed}  & UHSS                                 & $630$        & $704$      & $795$     & $6\%$         \\
            \cellcolor[HTML]{36E76A}                        & \textcolor{lightgray}{sc11}      &                &    & \textcolor{lightgray}{Mild}                                  & \textcolor{lightgray}{NaN}        & \textcolor{lightgray}{NaN}      & \textcolor{lightgray}{NaN}     & \textcolor{lightgray}{NaN}         \\
            \cellcolor[HTML]{A7F8C9}                        & sc12                                               & \multirow{-4}{*}{High}          & \multirow{-2}{*}{Pulsating} & UHSS                                 & $656$        & $746$      & $851$     & $12\%$         \\
            \bottomrule
        \end{tabular}
        \captionof{table}{Summary of \textit{simulation decomposition} of the structural reliability 4R model.}
        \label{tab:simdec}
    \end{tabularx}
\end{table}

Figure \ref{fig:simdec} demonstrates how non-monotonic the effects of the input variables on the model output are, and that the effect of one input is conditioned to the states of another one. First, the residual stress $\sigma_{res}$ divides the distribution into two narrow scenarios, \textit{medium} and \textit{high}, whereas the \textit{low} scenario sub-distribution spreads out through the entire range of the output values. Stress ratio $R$ has only a minor effect on the output in \textit{medium} and \textit{high} states of residual stress (small horizontal shift of the respective sub-distributions) but plays an important role in the \textit{low} residual stress, almost dividing the output into halves. In \textit{low} residual stress, steel grade $R_{p0.2}$ has a minor effect when the stress ratio is \textit{reversed} 
but creates a substantial rightward shift if the stress ratio is \textit{pulsating}. The combinations of \textit{medium} residual stress \& \textit{UHSS} steel grade and \textit{high} residual stress \& \textit{mild} steel grade are non-existent, which reflects physical material constraints and thereby explains the negative second-order effect. These visceral relationships would not have been revealed or apparent without the SimDec visualization.

\section{Conclusions}

This paper introduces and extensively tests the simple binning approach for computing first- and second-order sensitivity indices, which provide more accurate results than classic in global sensitivity analysis measures for Sobol' indices. The revealed conservation property of the combined indices in the presence of dependent inputs opens up a future research pathway of correlation-interaction intertwining in the second-order effects. Further, the paper demonstrates the importance of visualizing input-output relationships on a structural reliability model using a SimDec approach. The resulting methodological framework is streamlined (i.e., it works with a given data of as little as 1000 data points), open-sourced, and provides an intuitive way of 'looking' into a computational model and grasping its behavior. The presented framework can be recommended for analyzing disparate complex technological models for engineering design and decision-making.

\section*{Acknowledgements}
The authors are grateful to Prof. Art Owen for his comments in the early stages of this work. The work was supported by grant 220178 from the Finnish Foundation for Economic Education, and by grant OGP0155871 from the Natural Sciences and Engineering Research Council.



\bibliographystyle{johd}
\bibliography{bib}


\end{document}